\documentclass[]{spie}

\usepackage{amsmath,amsfonts,amssymb}
\usepackage{graphicx}
\usepackage[colorlinks=true, allcolors=blue]{hyperref}
\usepackage{todonotes}
\usepackage{subcaption}
\usepackage{color}
\usepackage{xcolor}
\usepackage[utf8]{inputenc}
\usepackage[T1]{fontenc}

\title{Far Sidelobes from Baffles and Telescope Support Structures in the Atacama Cosmology Telescope}

\author[a]{Patricio A. Gallardo}
\author[b]{Nicholas F. Cothard}
\author[c]{Roberto Puddu}

\author[d]{Rolando D\"unner}
\author[a]{Brian J. Koopman}
\author[a]{Michael D. Niemack}
\author[e]{Sara M. Simon}
\author[f]{Edward J. Wollack}

\affil[a]{Department of Physics, Cornell University, Ithaca, NY 14853 USA}
\affil[b]{Department of Applied and Engineering Physics, Cornell University, Ithaca, NY 14853 USA}
\affil[c]{Instituto de Astrof\'isica, Pontificia Universidad Cat\'olica de Chile, Macul, Santiago, Chile}

\affil[d]{Instituto de Astrof\'isica y Centro de Astroingenier\'ia, Facultad de F\'isica, Pontificia Universidad Cat\'olica de Chile, 7820436 Macul, Santiago, Chile}
\affil[e]{Department of Physics, University of Michigan, Ann Arbor, MI, 48103}
\affil[f]{NASA/Goddard Space Flight Center, Greenbelt, MD, USA}

\authorinfo{Further author information: (Send correspondence to PAG)\\PAG: E-mail: pag227@cornell.edu}

\begin{document} 
\maketitle

\begin{abstract}
The Atacama Cosmology Telescope (ACT) is a 6~m telescope located in the Atacama Desert, designed to measure the cosmic microwave background (CMB) with arcminute resolution. ACT, with its third generation polarization sensitive array, Advanced ACTPol, is being used to measure the anisotropies of the CMB in five frequency bands in large areas of the sky ($\sim 15,000$ $\rm deg^2$). These measurements are designed to characterize the large scale structure of the universe, test cosmological models and constrain the sum of the neutrino masses. As the sensitivity of these wide surveys increases, the control and validation of the far sidelobe response becomes increasingly important and is particularly challenging as multiple reflections, spillover, diffraction and scattering become difficult to model and characterize at the required levels. In this work, we present a ray trace model of the ACT upper structure which is used to describe much of the observed far sidelobe pattern. This model combines secondary mirror spillover measurements with a 3D CAD model based on photogrammetry measurements to simulate the beam of the camera and the comoving ground shield. This simulation shows qualitative agreement with physical optics tools and features observed in far sidelobe measurements. We present this method as an efficient first-order calculation that, although it does not capture all diffraction effects, informs interactions between the structural components of the telescope and the optical path, which can then be combined with more computationally intensive physical optics calculations. This method can be used to predict sidelobe patterns in the design stage of future optical systems such as the Simons Observatory, CCAT-prime, and CMB Stage IV.
\end{abstract}


\section{Introduction}
\label{sec:introduction}
\label{sec:intro}
The Atacama Cosmology Telescope\cite{swetz_atacama_2011} (ACT) is a 6 meter telescope intended to measure the cosmic microwave background (CMB) at arcminute scales and larger. Recent upgrades to the ACT have been deployed: ACTPol\cite{thornton_atacama_2016}, the first polarimeter consisting of three kilopixel cameras was deployed in 2013 and Advanced ACTPol (AdvACT)\cite{henderson_advanced_2016}, a set of three two-kilopixel cameras was deployed in 2016. These upgrades expanded the frequency coverage, increased the  sensitivity of the experiment\cite{naess_atacama_2014,louis_atacama_2017} and  were combined with enlarged  sky coverage\cite{de_bernardis_survey_2016}. Large area surveys motivate the need to understand the response of the system at large angles from the main beam as wider scans are more susceptible to sun and moon contamination. This contamination can be avoided if understood well, by incorporating the sidelobe map into the observation strategy design\cite{stevens_designs_2018} and by implementing baffles that control the sidelobes. 

Large angle sidelobes can be difficult to model and measure for large optical systems \cite{lockman_stray_2002,dunner_far_2012,fluxa_rojas_far_2016,barnes_first-year_2003,tauber_planck_2010}. They are usually generated by interactions between optical elements and non-idealities in the optical chain. Representative sources contributing to the telescope's sidelobe response include: multiple reflections between optical elements, reflections off baffling structures, scattering off optical elements and panel gap diffraction. Diffraction requires detailed electromagnetic modeling (see for example Flux\'a et al. 2016\cite{fluxa_rojas_far_2016}), which is often computationally intensive, making exploratory analyses impractical for systems that are thousands of wavelengths in size. In addition, it requires a detailed understanding of the material properties of all the internal components of the optical chain, some of which are difficult to obtain. Simpler techniques are needed in order to quantify the sidelobe pattern of the system.

In this work, we report a model that was built using commercial ray tracing software (Zemax OpticStudio\cite{zemax_llc_zemax_2016}) to simulate the far sidelobe pattern of the Atacama Cosmology Telescope (ACT) based upon the measured near field power pattern $P(\theta)$ of the receiver camera. This model combines measurements of the geometry of the telescope from photogrammetry with a beam model fit from measurements of the receiver camera sidelobe pattern as it was installed in the field. This modeling technique allows rapid prototyping (with run times of fractions of hour for a quad core machine) as it is able to identify the dominant part of the sidelobe pattern after a finite number of reflections. Polarization can be included in this model, but it is not explored here and is left as a future work. This is important as in general scattering and reflections can induce polarization\cite{renbarger_measurements_1998}.  This model explains a number of sidelobe features observed in ACT far sidelobe maps, we also show general agreement with a partial physical optics model which is under development (with run times of $\sim 30$ hours in a 32-core machine).

We compare the measured camera beam with physical optics models of the receiver camera and find that the measured camera spillover level is inconsistent with diffraction alone (for perfectly absorbing camera inner walls) at the Lyot stop and camera lenses. We discuss how this relates to the existence of scattering inside the camera and current work in progress to understand and improve this for future CMB experiments. This modeling technique is currently being used to inform the design stage of the Simons Observatory and CCAT-prime. In Section \ref{sec:photogrammetry}, we discuss how the 3D CAD model was built from measurements performed with  photogrammetry. Section \ref{sec:beam_mapping} shows the experimental setup to measure the camera beam map. Section \ref{sec:ray_trace} discusses the optical ray trace model and Section \ref{sec:results_and_discussion} results. 

\section{Photogrammetry model}
\label{sec:photogrammetry}
\label{sec:photogrammetry}
We built a 3D CAD model that describes the geometry of the upper structure of ACT. This model includes the two mirror geometry, comoving ground shield, secondary mirror baffle and camera baffle. The positions of each panel in the comoving shield and baffling structure were measured with a photogrammetry system to describe the geometry of the telescope in its current state after minor modifications throughout the seasons.

\begin{figure}
\centering
\includegraphics[width=0.7\textwidth]{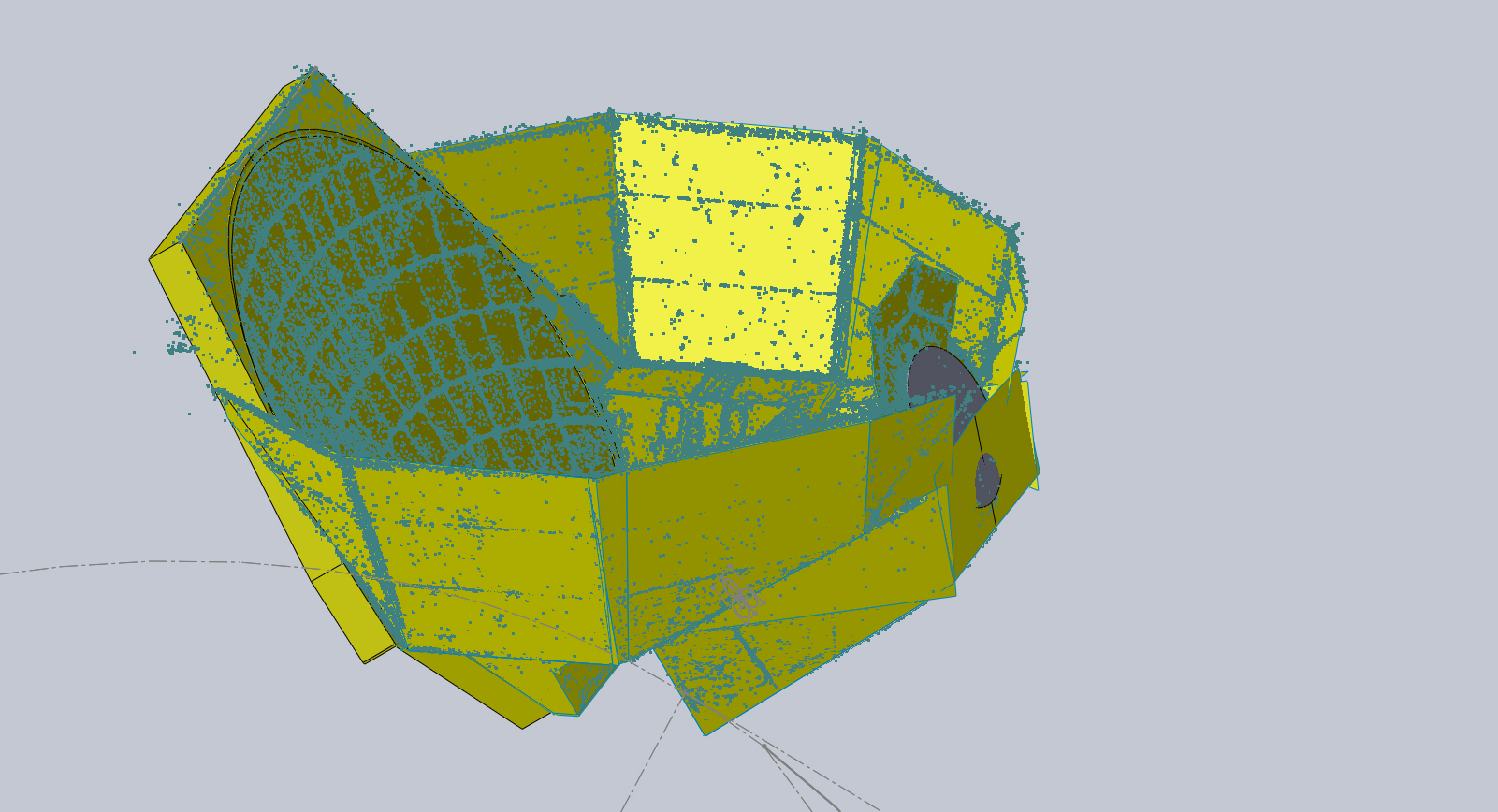}

\caption{The point cloud obtained with photogrammetry via Agisoft Photoscan, and the CAD model of the telescope built from the point cloud. Primary and secondary mirrors follow the optical design of the telescope\cite{fowler_optical_2007}. The paneled secondary guard (flat panels around the secondary mirror of which the top part is shown) ring and the comoving ground shield are shown. This model was imported as a solid model to Zemax for the ray trace.}
\label{fig:CAD_model}
\end{figure}

The photogrammetry system consists of a series of patterned stickers (photogrammetry targets) which are photographed by a Nikon 600D camera and processed with the Agisoft Photoscan\footnote{http://www.agisoft.com/} software that generates the 3D positions of these targets. We use the measured positions of the targets to create a best fit model that represents the shape of the telescope upper structure's planar faces. The primary and secondary mirrors are drawn using the optical specifications of ACT\cite{fowler_optical_2007}. Figure \ref{fig:CAD_model} shows the resulting 3D model. This CAD model was drawn in SolidWorks\footnote{http://www.solidworks.com}, and it can be exported for use in Zemax.

\section{Beam Mapping and Camera Beam Model}
\label{sec:beam_mapping}
\label{sec:camerabeam} 

\begin{figure}

\centering
	\begin{subfigure}{0.48\textwidth}
	\includegraphics[width=\textwidth]{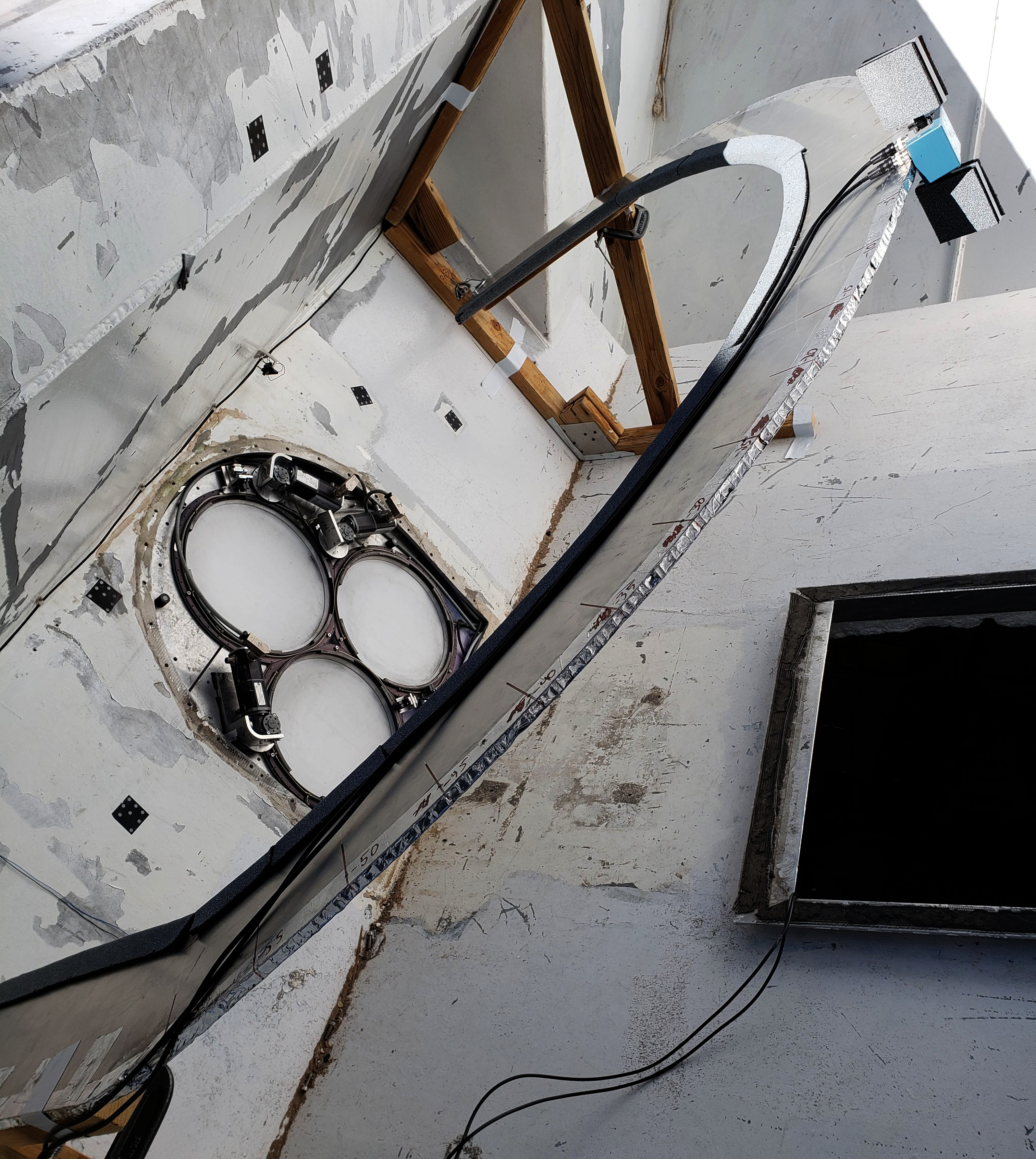}
    \label{subfig:beam_mapperSetup}
	\end{subfigure}
    \begin{subfigure}{0.48\textwidth}
    \includegraphics[width=\textwidth]{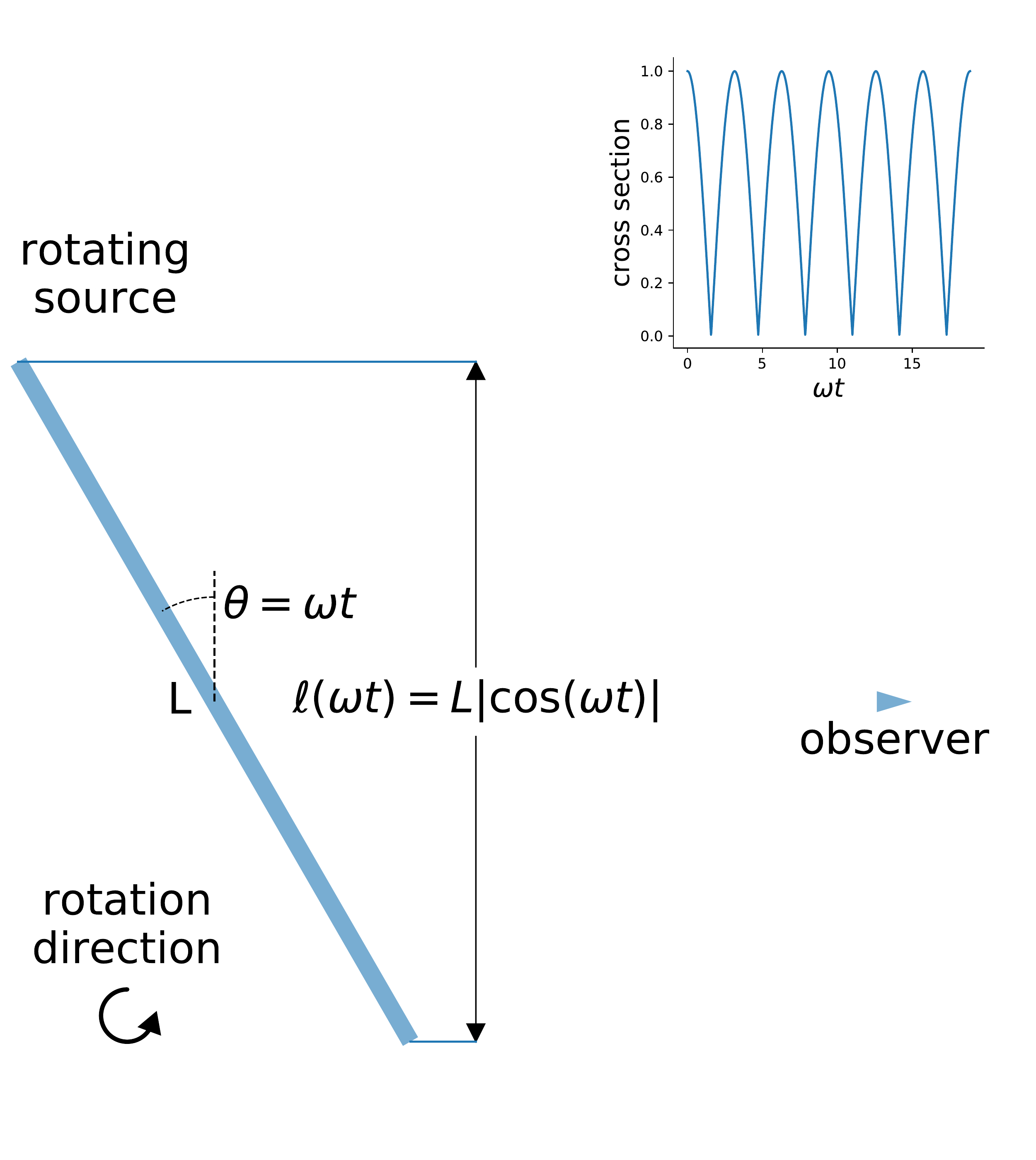}
    \label{subfig:beamMapperSetup}
    \end{subfigure}
\caption{Left: Beam mapping setup mounted in front of the AdvACT's mid frequency (MF2) optics tube window. A semicircular arch (radius $\sim1.25$ m) is centered and normal to the window. The stepper motor-modulated Eccosorb source is mounted on the arch at the $0^\circ$ position. The source can easily be placed at any $5^\circ$ increment along the arch between $-60^\circ$ and $+60^\circ$. Right: Diagram describing the setup. A rotating source presents an area to the observer that depends on its angle $\theta$ with respect to the arch. The cross section ($A=\ell \times L$ for a square) is proportional to the projection of one of the sides of the square $\ell = L |\cos{\omega t}|$. Upper right shows the expected shape of the waveform for an infinitely thin source.}
\label{fig:beam_mapper_setup}
\end{figure}

We measured the optical power detected in the AdvACT mid-frequency (MF2) detector array for different angles\cite{gallardo_mirror_2012} with an ambient temperature source in front of the camera. Optical loading was varied by spinning $7.5~\rm{cm} \times7.5~\rm{cm}$ squares of AN-72 Eccosorb\footnote{http://www.eccosorb.com/} at specific angles in front of the MF2 optics tube window (aperture ID: $\sim 30\, \rm cm$). Accurate positioning of the Eccosorb was achieved using a semicircular ($\sim1.25$ m in radius) hexcell arch mounted horizontally in front of the MF2 window. Angular intervals every five degrees from the boresight were marked on the arch, providing a ruler from $-60^\circ$ to $+60^\circ$. For each angular position, the Eccosorb squares were spun at 0.8 Hz for a time interval of three minutes using a stepper motor (200 steps per full rotation) that was mounted to the hexcell arch at each five degree increment. The cross-section of the Eccosorb, as seen by the camera, varies at a frequency of 1.6 Hz (as shown in the right panel of Figure \ref{fig:beam_mapper_setup}). Figure \ref{fig:beam_mapper_setup} (left) shows the hexcell arch mounted in front of MF2, with the stepper motor and Eccosorb squares mounted at the $0^\circ$ position.

\begin{figure}
\centering
\includegraphics[width=0.9\textwidth]{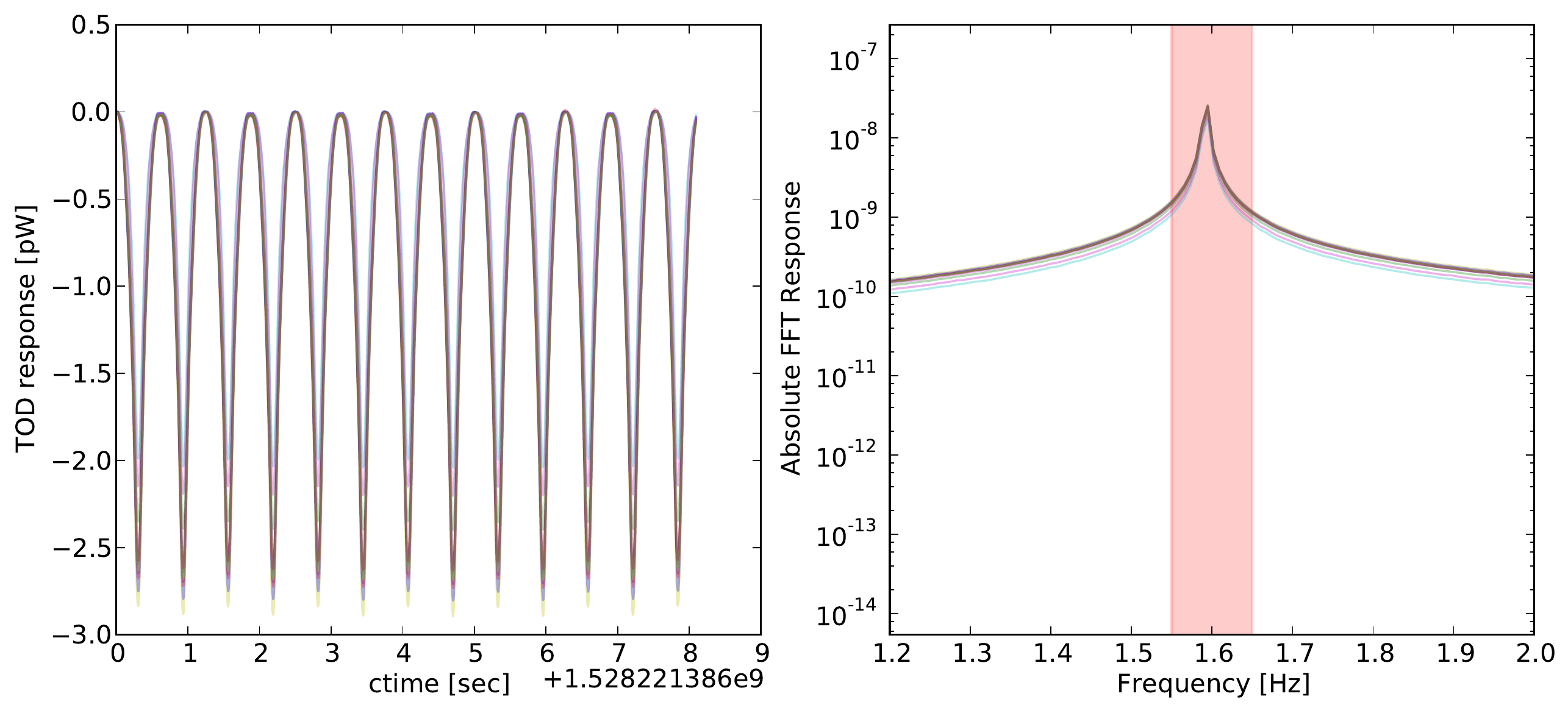}
\caption{Detector time streams and their Fast Fourier Transform (FFT) while the source was placed at $0^\circ$ on the arch. Each line represents the time stream and FFT of a single $150 \,\rm GHz$ detector. Left: 8 second clip of a 180 second time stream as measured by the $150\,\rm GHz$ detectors within $10\, \rm mm$ of the center of the MF2 array. The modulated signal is clear and gives the expected $f(t)=|\cos(\omega t)|$ response. Right: FFT of one 180 second time stream clearly showing power at $1.6 \, \rm Hz$. The $\sim 2.5\,\rm{pW}$ amplitude of the signal is a significant fraction of the $\sim 10\,\rm{pW}$ saturation power of these bolometers. Preliminary analysis suggests that this may lead to modest compression of the main beam at $150 \, \rm GHz$, which we are continuing to study.}
\label{fig:beam_mapper_TOD}
\end{figure}

\begin{figure}
\centering
\includegraphics[width=0.9\textwidth]{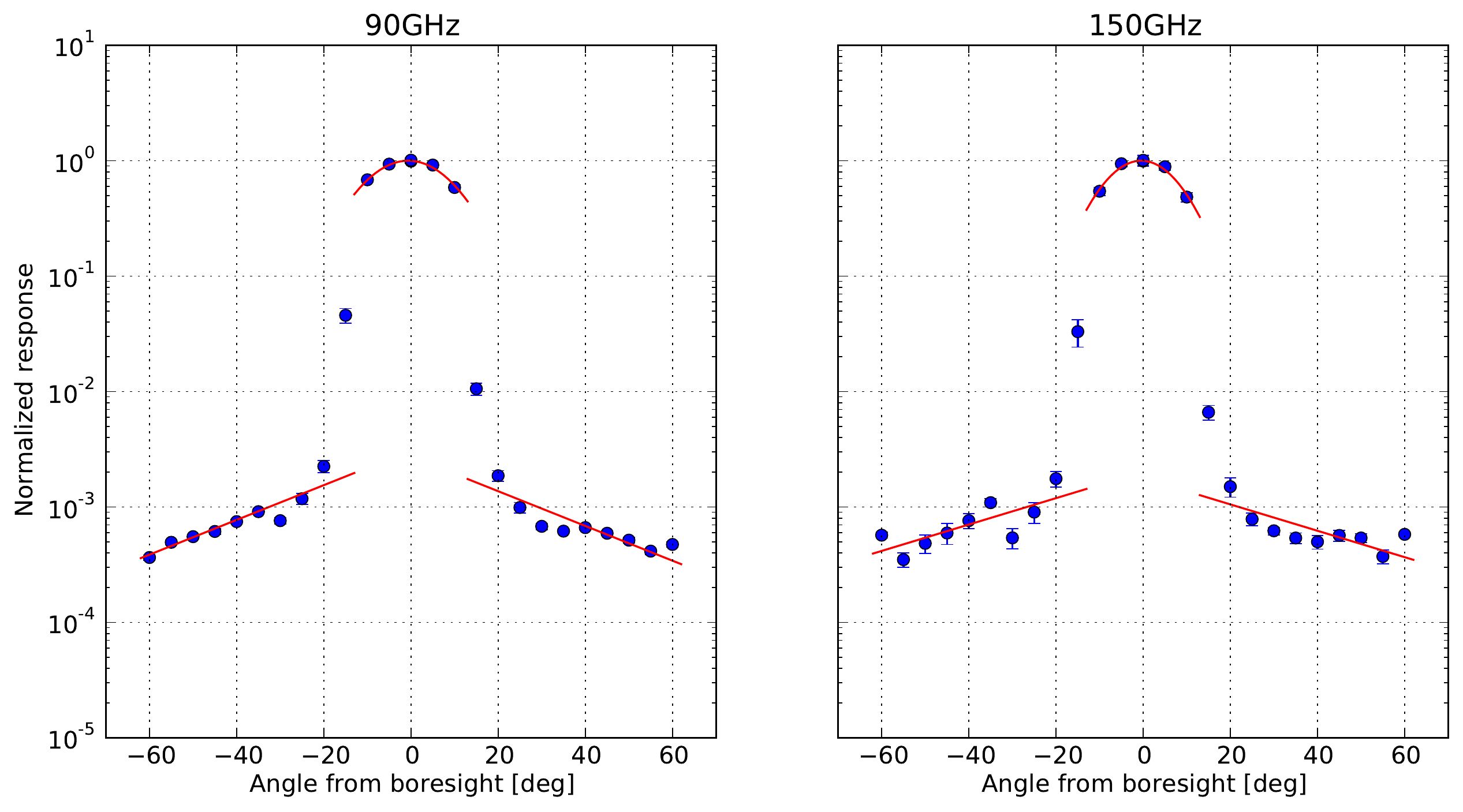}
\caption{Camera beam map for 90 and 150 GHz detectors of the MF2 array. The data fits to a Gaussian main beam and exponential sidelobes. The cutoff angle between the Gaussian and exponential components is set to $11.3^\circ$ as determined by the Lyot stop diameter in the optics tube. The encircled energy outside the main beam is 3.1\% and 3.3\% for 90 and 150 GHz respectively. Excess loading from the modulator may have caused modest compression at the central angles which may artificially increase the relative power of the sidelobes. These measurements are expected to set an upper limit on the sidelobe power.}
\label{fig:beam_mapper_map}
\end{figure}

The modulation of the Eccosorb cross section was measured by the feedhorn coupled transition-edge sensors (TES) in the MF2 detector array. An example of the time stream data is shown in Figure \ref{fig:beam_mapper_TOD} (left). FFTs of the detector time streams show signal in a 1.6 Hz tone as shown in Figure \ref{fig:beam_mapper_TOD} (right). Only detectors within $10\, \rm mm$ of the array center were used in the analysis because their beams are expected to be centered on the cryostat window. By comparing the power in the $1.6\, \rm Hz$ tone at each angle, an estimate of the relative optical intensity was obtained. The power in the $1.6\, \rm Hz$ tone was computed at each angle, by integrating the power spectral density from 1.55 to 1.65 Hz for each detector. The camera beam power pattern amplitude for a particular angle is taken to be the mean integral of the power spectral density over the mentioned band and over the sampled detectors. The uncertainty on this method is estimated as the standard deviation of the integral over the sampled detectors. Work is ongoing to improve this estimate by considering systematic effects, such as the source physical size or precision in the source placement. Excess loading while the modulator was near central angles may have driven the detectors into non-linear regimes and is being investigated. A non-linear response would cause some compression of the modulated signal, resulting in a under-estimated beam response at central angles. This effect could artificially increase the relative power in the sidelobes and so these measurements currently set an upper limit on the sidelobe power. Figure \ref{fig:beam_mapper_map} shows the measured beam profile of the $90\, \rm GHz$ and $150\, \rm GHz$ bands in MF2. 

We fit the beam pattern to a piecewise functional form described by 
\begin{equation} \label{eq:beam_mapper_eqs}
\text{CameraBeam}(\theta)=
\begin{cases}
 & \exp\left( \frac{-(\theta-c_g)^2}{s_g^2} \right )   \text{ for } \theta < 11.3^\circ \\ 
 & A_e\exp\left( \frac{-|\theta-c_e|  }{s_e  } \right )   \text{ for } \theta > 11.3^\circ ,
\end{cases}
\end{equation}
which is informed by the design of the instrument. This model contains a Gaussian central beam (produced by the feedhorns in front of the detector array) with a width parameter given by $s_g$ truncated at an angle ($\theta=11.3^\circ$) given by the diameter of the Lyot stop. The exponential fall-off is found phenomenologically and its overall amplitude is given by $A_e$, while the speed of the fall-off is given by $s_e$. Finer angular structure is not captured in this model as its main purpose is to measure the overall level.

In Figure \ref{fig:beam_mapper_map}, the solid lines show the best fit for a Gaussian main beam and exponential sidelobes. The best fit parameters for Equation \ref{eq:beam_mapper_eqs} are given in Table \ref{tab:beamParams}. The encircled energy outside the central Gaussian beam is 3.1\% for 90 GHz and 3.3\% for 150 GHz.

\begin{table}[h]
\centering
\begin{tabular}{l|ccccccc}
Parameter &  $s_g [\rm deg]$ & $c_g [\rm deg]$ & $A_e [\rm -]$ & $s_e [\rm deg]$ & $c_e [\rm deg]$  \\\hline\hline
90 GHz    & 15.1 & -0.6 & $2.91\times10^{-3}$ & 28.9 & -1.7 \\
150 GHz   & 12.7 & -0.4 & $1.90\times10^{-3}$ & 38.0 & -2.4 \\
\end{tabular}
\caption{Best fit parameters of the beam map data shown in Figure \ref{fig:beam_mapper_map}. Parameters are as defined in Equation \ref{eq:beam_mapper_eqs}.}
\label{tab:beamParams}
\end{table}

\section{Ray trace model}
\label{sec:ray_trace}
\label{sec:raytracemodel}

The beam measured in section \ref{sec:camerabeam} represents the near field power pattern $P(\theta)$ of the receiver camera as seen from the secondary with origin at the center of the camera. The far field limit, given as $d_{ff} = \frac{2 D^2}{\lambda}$, where D is the diameter of the aperture in question (in this case $\sim 30 \, \rm cm$) and $\lambda$ is the wavelength of observation ($3.3$ or $2\, \rm mm$)   occurs at $55$ or $90 \,\rm m$ from the camera aperture (at $90$  or $ 150 \, \rm GHz$).

The fit model of $P(\theta)$ is used to launch rays from the camera window. The probability of launching a ray in a certain direction is proportional to the amplitude of the fit power pattern. A tilt was applied to the forward direction of each beam depending on the optics tube that was being modeled. This tilt is given by the chief ray director cosines from the optical design of AdvACT. The power pattern is assumed to be rotationally symmetric, have a Gaussian dependence for angles where the camera illuminates the secondary (in the ray trace limit) and have an exponential decay for large angles as described in Section \ref{sec:camerabeam}.

\begin{figure}
\centering
	\begin{subfigure}{0.48\textwidth}
	\includegraphics[height=0.8\textwidth]{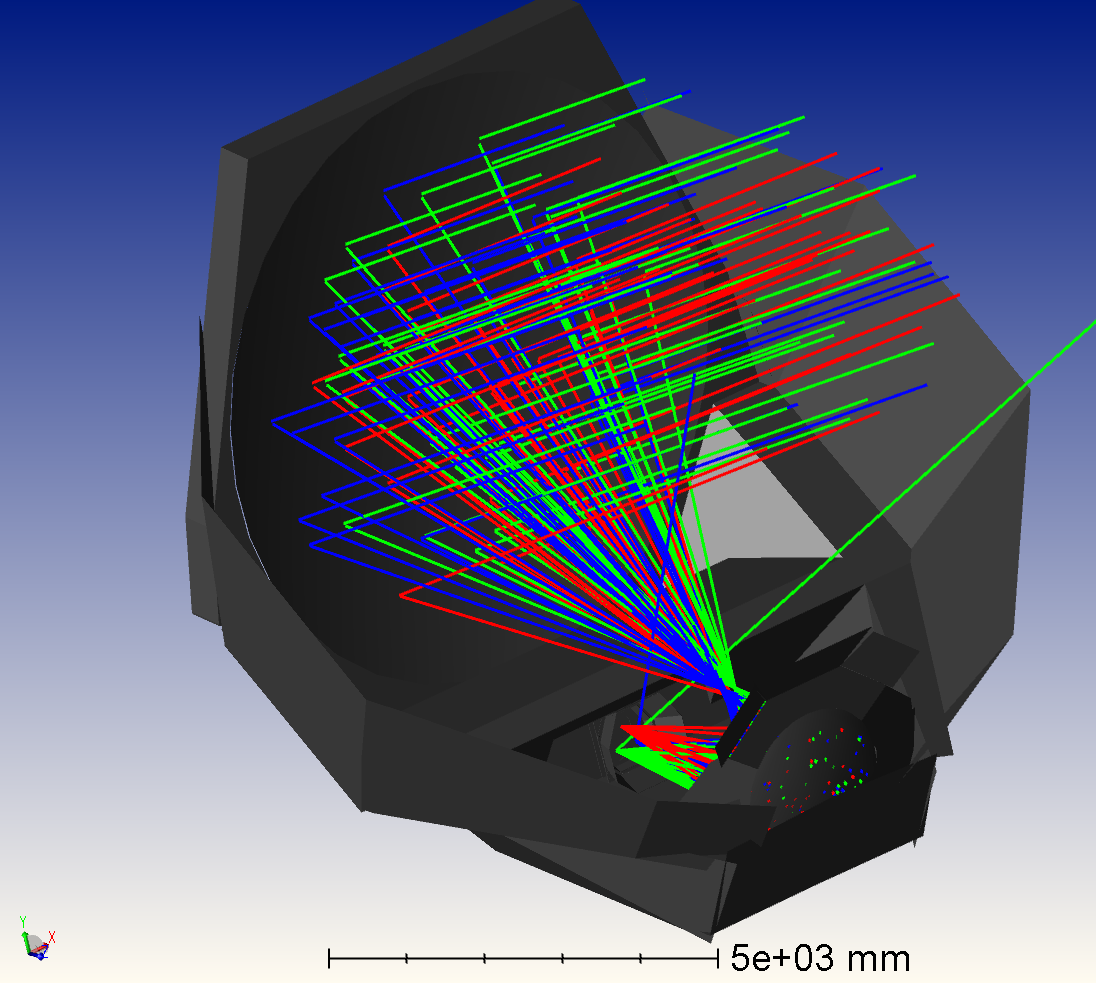}
	\caption{}
    \label{subfig:raytrace}
	\end{subfigure}
    \begin{subfigure}{0.48\textwidth}
    \includegraphics[height=0.8\textwidth]{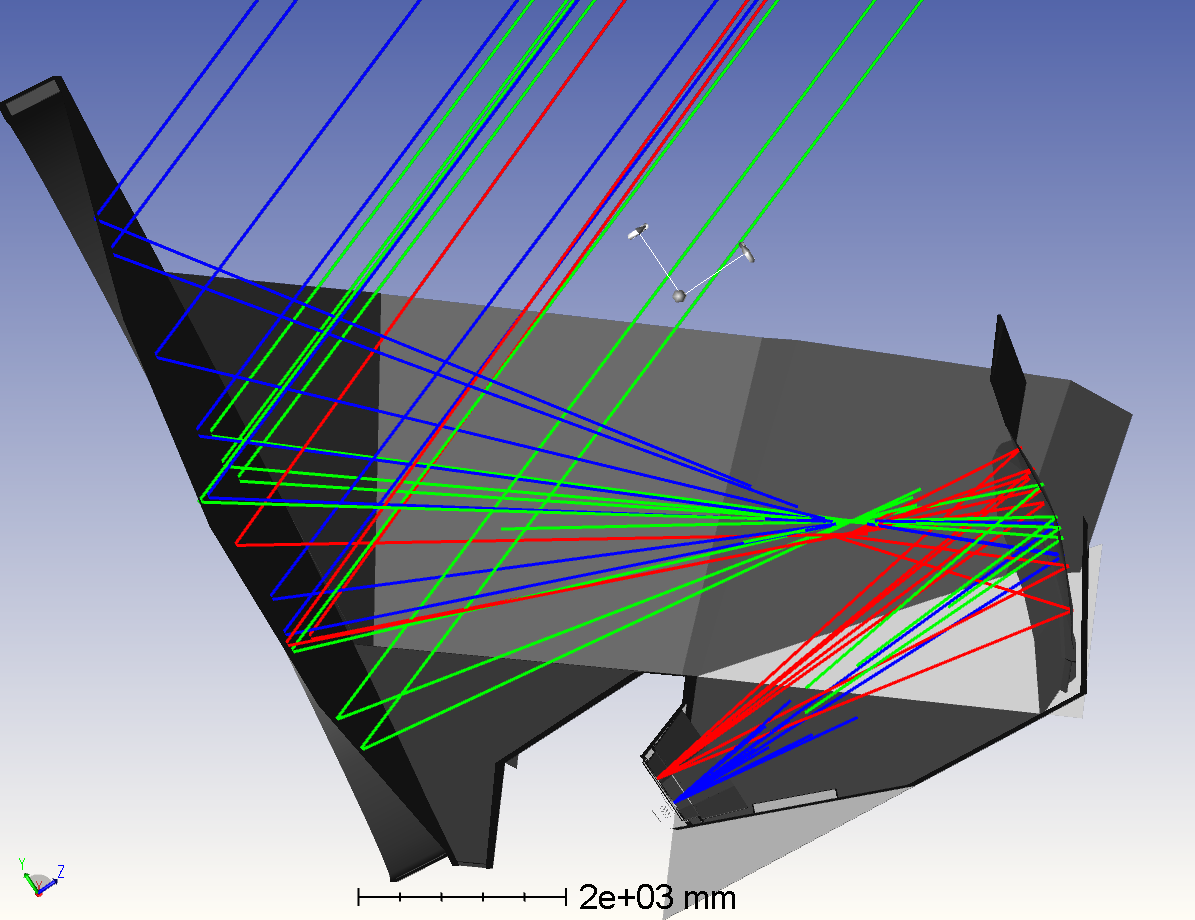}
    \caption{}
    \label{subfig:raytrace_cut}
    \end{subfigure}
\caption{Solid model of ACT with one example ray trace of 30 rays per field. Three fields (green, red and blue) are shown, corresponding to the center of the three cameras in the ACT receiver. In this example, two stray rays can be seen (in green and blue), showing the direct line of sight sidelobe from the camera to the sky. Multiple bounce sidelobes emerge after a large number of rays are traced.}
\label{fig:zemaxModel}
\end{figure}

In Zemax, we place a source with the best fit beam $P(\theta)$ at the center of one of the camera windows (PA3) to simulate the response of this camera to off-axis radiation in the time-reverse sense. Figure \ref{fig:zemaxModel} shows an example of ray traces for a limited number of rays. We launch $10^7$  rays to sample the sidelobe pattern far from the main beam in this solid model and count the rays in an spherical grid surrounding the telescope. Total power injected in this simulated system from the focal plane is 1 Watt, which is used as normalization for the angular power density.

\section{Results and discussion}
\label{sec:results}
\label{sec:results_and_discussion}

\begin{figure}
\centering
	\begin{subfigure}{0.48\textwidth}
		\includegraphics[width=\textwidth]{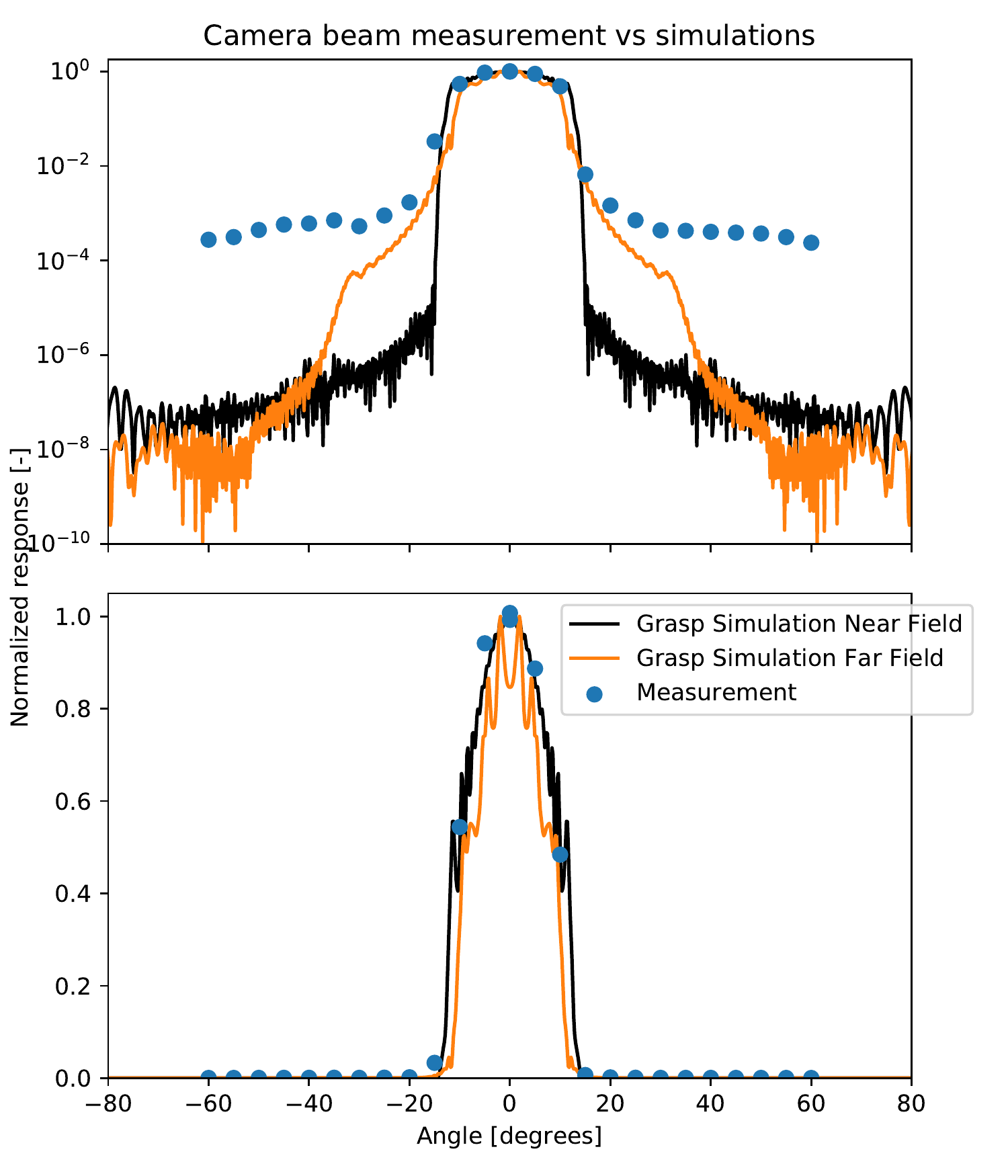}
    \end{subfigure}
    \begin{subfigure}{0.48\textwidth}
    	\includegraphics[width=\textwidth]{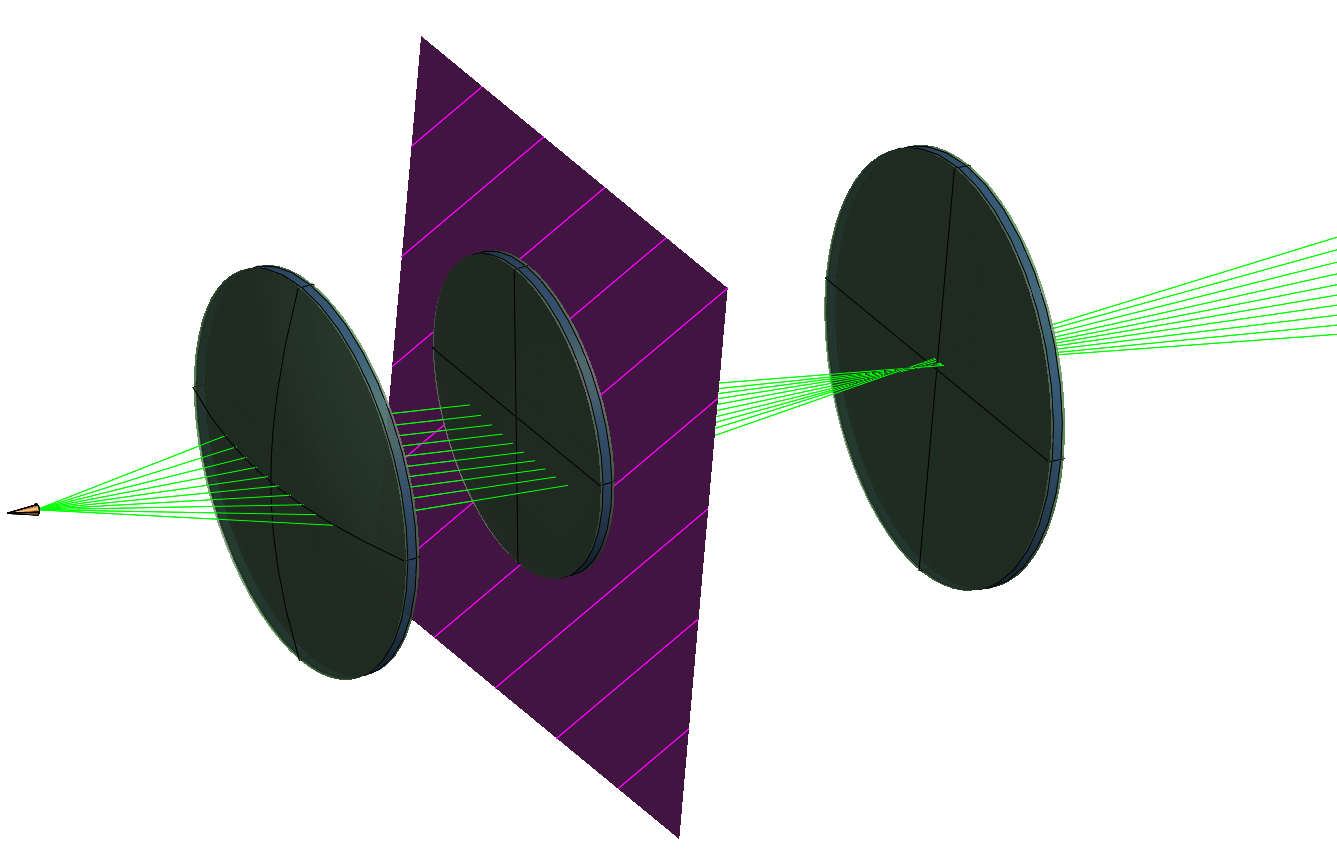}
    \end{subfigure}
    
\caption{Left: Measurement and simulations of the AdvACT camera beam. Measurements show a level of sidelobes higher than what is expected from physical optics simulations of a perfectly absorbing wall camera interior model. Preliminary ray tracing simulations in Zemax with a scattering function at the camera inner walls are able to describe the discrepancy in the level of scattering. Detailed measurements of these scattering functions can be used to refine this model. Right: Schematic of GRASP model (three lens and Lyot stop) used to produce the physical optics beams shown on the left.}
\label{fig:GraspVsMeas}
\end{figure}

The measured level of sidelobes for the camera beam is significantly higher than what is expected from a physical optics diffraction only model (GRASP\footnote{http://www.ticra.com}) with perfectly absorptive walls (shown in Figure \ref{fig:GraspVsMeas}). It should be noted that the measured camera beam is expected to set an upper limit on the sidelobe power due to possible compression of the main beam. Figure \ref{fig:GraspVsMeas} shows a linear and log comparison of the measured data points with the spillover level at large angles. This discrepancy is consistent with scattering being the dominant source in the camera sidelobe pattern. Measurements of the reflectance of the blackening agents used during the construction of the camera are under way and are being considered for the construction of the Simons Observatory.

\begin{figure}
\centering
	\begin{subfigure}{0.48\textwidth}
	\includegraphics[width=\textwidth]{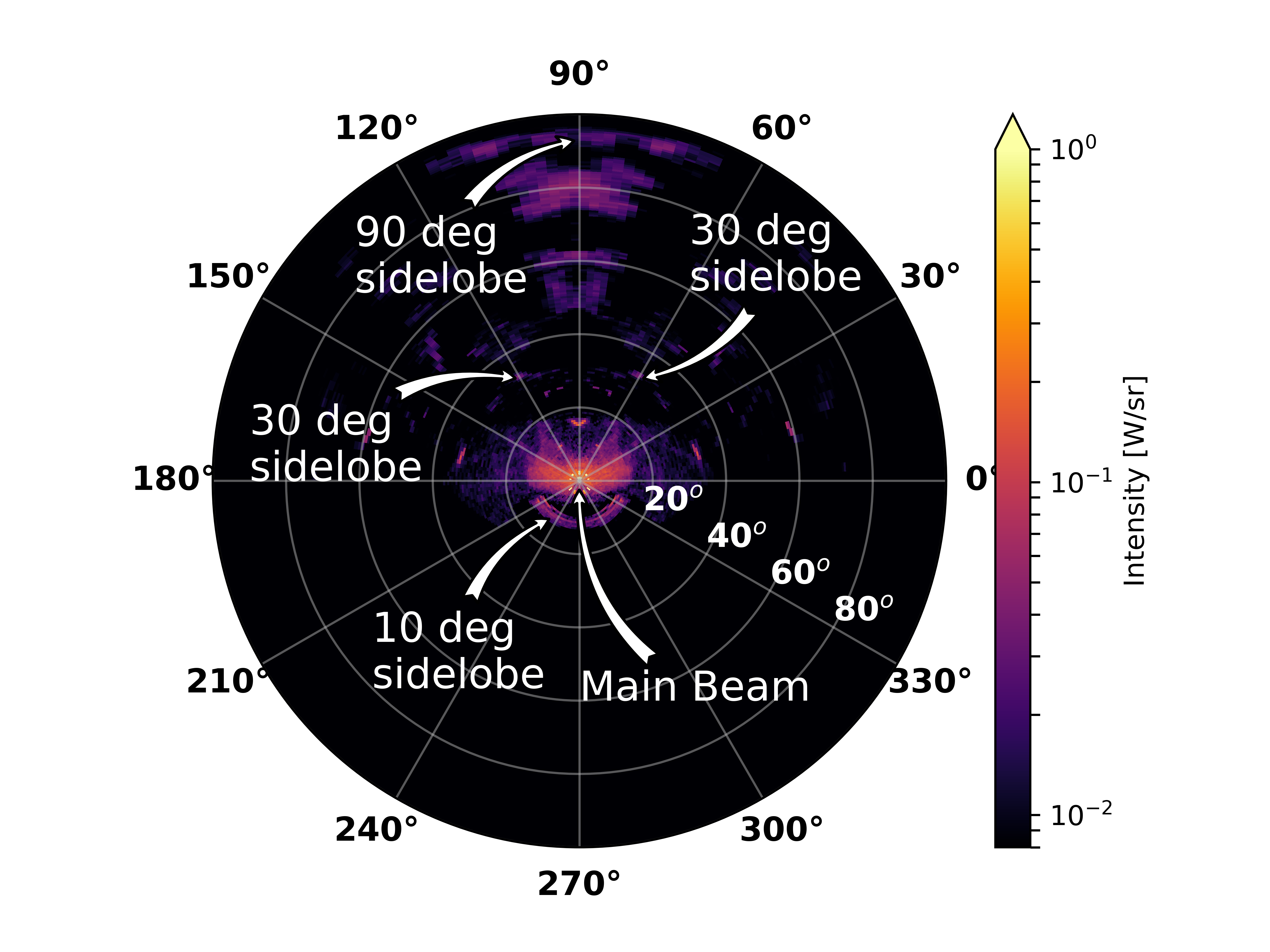}
	\end{subfigure}
    \begin{subfigure}{0.48\textwidth}
	\includegraphics[width=\textwidth]{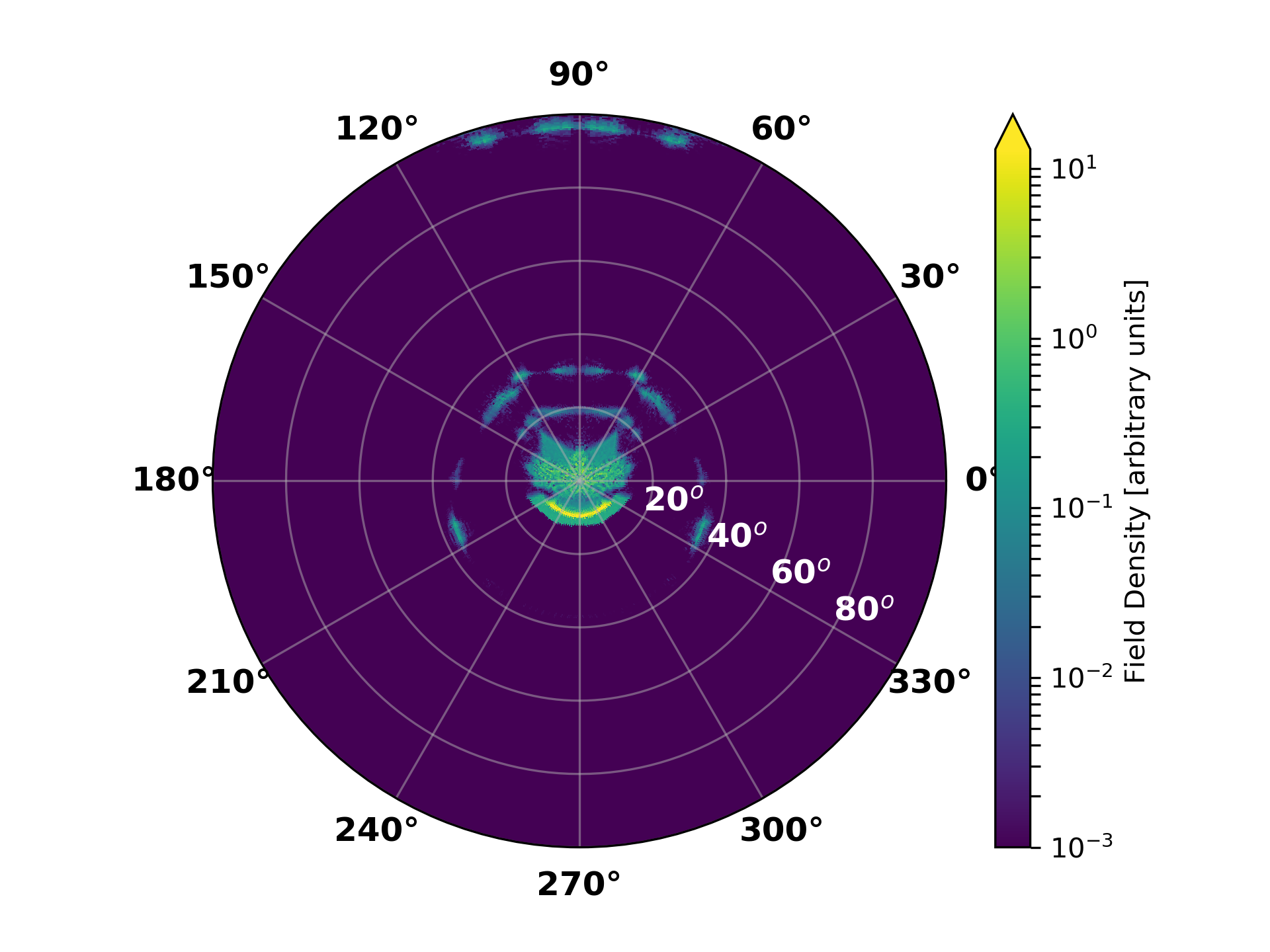}
	\end{subfigure}
    
\caption{Left: Sidelobe map generated with the ray trace presented here. This is the result of one ray trace of $1\times 10^7$ rays. Arrows show positions of known far sidelobes. The $10^\circ$ sidelobe is present under the main beam and has an arc shape. The $30^\circ$ sidelobe has mirror symmetry for the center camera, this symmetry is broken when the simulation is run for different pixels in the camera window, however the arc shape is preserved. Right: Grasp simulation of a partial model of ACT. The simulation includes primary mirror, secondary baffle and primary baffle. Secondary mirror is omitted here so the sidelobes can be seen. The comoving structure is missing in this simulation, which produces artifacts in the lower part of the map (like the features near the $40^\circ$ mark near $330^\circ$ in the azimuthal coordinate).}
\label{fig:sidelobeMaps}
\end{figure}

There is qualitative agreement between the ray trace model presented here and moon maps made during observations. This simulation also agrees qualitatively with a partial (primary, secondary baffle and primary guard ring, without comoving structure) telescope model physical optics calculations done in Grasp. Comparison is shown in Figure \ref{fig:sidelobeMaps}. Common features between this model, the physical optics calculation and the  and far sidelobe moon maps include the  sidelobes at 10, 30 and $90^\circ$ shown in Figure \ref{fig:sidelobeMaps}. We have addressed lunar pickup through a combination of adjusting the survey strategy to increase moon avoidance and cutting data when the moon is located in the dominant sidelobes. The overall level of the sidelobes can be difficult to extract from this simulation as diffraction is not included. However, through a combination of comparing the beam maps and simulations with planet beam measurements\cite{thornton_atacama_2016} we find that the peak sidelobe amplitude in the simulations is approximately $60 \, \rm dB$ below the main beam. This is consistent with estimates of moon pickup in far sidelobe measurements\cite{naess_atacama_2014}. Further study is needed to contrast this with experiments and physical optics simulations.

\section{Conclusion}
We present a method for rapidly evaluating the sidelobe pattern of a millimeter wave experiment. This method requires as an input the camera near field power pattern, for which we discuss measurements recently made in Advanced ACTPol. We find qualitative agreement between the observed sidelobe pattern (from moon maps) and the output of the angular distribution of scattered light by this model. This method is able to explain what surfaces in the structure of the telescope are responsible for the features observed in the sidelobe map. This method is fast and can be run in a desktop computer allowing exploratory analysis of baffling strategies that minimize the impact of sidelobes in the data. This modeling approach is being used to control sidelobes and inform the design of new telescopes for the Simons Observatory and CCAT-prime. Understanding and controlling these effects will become increasingly important as new observatories like these extend CMB measurements to much lower noise levels.

\acknowledgments 
The ACT project is supported by the U.S. National Science Foundation through awards AST-1440226, AST- 0965625 and AST-0408698, as well as awards PHY-1214379 and PHY-0855887.  MDN and PAG acknowledge support from NSF award AST-1454881. Work by NFC was supported by a NASA Space Technology Research Fellowship. R.D. thanks CONICYT for grants PIA Anillo ACT-1417 and QUIMAL 160009.


\bibliography{Zotero}
\bibliographystyle{spiebib}

\end{document}